\documentclass{article}

\usepackage{graphicx}
 \usepackage{mathptmx}      
\usepackage[mathscr]{eucal}
\usepackage{amsmath,amssymb}
\usepackage{graphicx}
\usepackage{subfig}
\usepackage{color}

\def \beq{ \begin{equation} }
\def \eeq{\end{equation}}

\DeclareSymbolFont{AMSb}{U}{msb}{m}{n}
\DeclareMathSymbol{\bdC}{\mathbin}{AMSb}{'103}
\DeclareMathSymbol{\bdR}{\mathbin}{AMSb}{'122}
\DeclareMathSymbol{\bdN}{\mathbin}{AMSb}{'116}

\begin{document}

\title{Central Configurations of the Five-Body Problem with Equal Masses}

\maketitle
\author{\begin{center}
Tsung-Lin Lee\\
\smallskip
{\footnotesize Department of Mathematics\\
 Michigan State University\\ East Lansing, MI 48824, USA\\
leetsung@msu.edu\\
}\end{center}

\begin{center}
Manuele Santoprete\\
\smallskip
{\footnotesize Department of Mathematics\\
Wilfrid Laurier University\\
75 University Avenue West,\\
Waterloo, ON, Canada, N2L 3C5.\\
msantopr@wlu.ca\\
}\end{center}
}






\begin{abstract}
In this paper we present a complete classification
of the isolated central configurations of the five-body problem with
equal masses.
This is accomplished by using the polyhedral homotopy method to
approximate all the isolated solutions of the Albouy-Chenciner
equations.
The existence of exact solutions, in a neighborhood of the
approximated ones, is then verified using the Krawczyk method.
Although the Albouy-Chenciner equations for the five-body problem
are huge, it is possible to solve them in a reasonable amount
of time.
%
\end{abstract}


\section{Introduction}

The Newtonian $n$-body problem is the study of the dynamics of $n$
point particles with masses $m_i\in{\mathbb R}^+$ and positions
$q_i\in{\mathbb R}^d$ ($i=1,\ldots,n$), moving according to
Newton's laws of motion:
\begin{equation}
m_j\ddot q_j=\sum_{i\neq j}\frac{m_im_j(q_i-q_j)}{r_{ij}^3}\quad 1\leq j\leq
n
\end{equation}
where $r_{ij}=\|q_i-q_j\|$ is the distance between $q_i$ and $q_j$.

In the Newtonian $n$-body problem, the simplest possible motions are
such that the configuration is constant up to rotations and scaling,
and each body describes a Keplerian orbit. Only some special
configurations of particles are allowed in such motions. Wintner
called them {\itshape central configurations}. A configuration
$(q_1,\ldots, q_n)$ is called a central configuration if and only if
there exists a $\lambda\in{\mathbb R}$ such that

\begin{equation}
\lambda(q_j-q_G)=\sum_{i\neq j} \frac{m_i(q_i-q_j)}{r_{ij}^3} \quad 1\leq
j\leq n  \label{cc}
\end{equation}
where $q_G=\sum_i m_i q_i / \sum_i m_i$ is the center of mass.
Equations (\ref{cc}) are invariant under rotations, dilatations and
 translations on the plane. Two central
configurations are considered equivalent if they are related by
these symmetry operations.

The question of the existence and classification of central configuration is
a fascinating problem that dates back to the 18th century. In 1767, Euler
discovered the collinear c.c.'s. In 1772 Lagrange proved that, for any three
arbitrary masses, the equilateral triangle is a central configuration.

For the collinear $n$-body problem an exact count of the central
configurations of $n$ bodies was found by Moulton \cite{Moulton} (see also
\cite{smale70} for a modern proof). There is a unique collinear relative
equilibrium for any ordering of the masses so there are $n!/2$ collinear
equivalence classes.

The number of planar central configurations of the $n$--body problem
for an arbitrary given set of positive masses, has been established
only for $n=3$: there are always five relative equilibria. Two of
these are Lagrange's equilateral triangles and the other three are
collinear c.c. discovered by Euler. Already in the four-body problem
there is sufficient complexity to prevent a complete classification
of the non-collinear relative equilibria. In fact, an exact count is
known only for the equal masses case \cite {Albouy95,Albouy96} and
for certain cases where some of the masses are assumed sufficiently
small \cite{Xia91,Tien93}. Some partial result in the four-body
problem with some equal masses are given in \cite
{Long,Santoprete07,Albouy08}.

Even the finiteness of the central configurations is a very difficult
question. This conjecture was proposed by Chazy \cite{Chazy} and Wintner
\cite{Wintner} and was listed by Smale as problem number $6$ on his list of
problems for this century \cite{Smale98}. The finiteness problem was settled
by  Albouy \cite{Albouy95,Albouy96} for the case of four equal masses and
by Hampton and Moeckel \cite{Hampton06} for the general four
body problem.

Aside from these fundamental results very little else is known in
terms of the classification of planar c.c.'s for $n\geq 4$. Strictly
spatial five-body central configurations are analogous in some ways
to the planar four-body case, and a classification of the symmetric
ones is given in \cite{Kotsireas}. Planar five-body central
configurations present new challenges and there are very few results
in this area.

In this paper, we present all the isolated central configurations of
the five-body problem with equal masses. Such central configurations
have been obtained by finding all the isolated solutions of the
Albouy-Chenciner equations (see Section \ref{AC} below, and
\cite{Albouy1998,Hampton06}) for this problem. Numerical explorations
of the problem have been conducted by Moeckel \cite{Moeckel89} and Ferrario
\cite{Ferrario}. They obtained isolated solutions using root-finding
routines with a random starting point. However, unfortunately, those
numerical experiments do not guarantee that there are no other
solutions, and do not ensure that approximate solutions correspond
to exact solutions.
Our approach is different, and it is based on the homotopy
continuation method.
This is a technique that was proposed to find  \textit{all} the
isolated solutions of polynomial systems \cite{HS,Lihand}.
We used the \textit{polyhedral} homotopy method to find approximate
solutions because it needs to trace fewer curves than other homotopy
methods.

Already for four bodies the Albouy-Chenciner equations are quite
large -- $12$ equations in $12$ variables with total degree
$2,985,984$. ~Moreover, even when the polyhedral homotopy method is
applied, $82,593$ curves need to be traced. Experiments running on a
3.2 GHz CPU machine show that the polynomial system solver
HOM4PS-2.0 obtains all the isolated solutions of the four-body
problem in 10 minutes, while other similar software solvers require
more than 10 hours to perform the same task. 
Considering that, in the five-body case,  the number of curves that need to be traced reaches $439,690,761$~ we use the kernel of HOM4PS-2.0 to find approximated isolated solutions. 
We then verify the existence of an exact solution in
a small neighborhood of the approximate one by using  the Matlab toolbox
INTLAB, an implementation of the interval Krawczyk method
~\cite{Kraw,Moore77}.

Note that, while the polyhedral homotopy method finds all the
isolated complex solutions, this does not exclude the existence of
positive-dimensional components of the variety defined by the
Albouy-Chenciner system in the algebraic torus $({\mathbb C}^*)^{20}=({\mathbb C}\setminus\{0\})^{20}$. Hence, at
least in principle,  there could still conceivably be undiscovered
``continua'' of positive real solutions. The fact that this is a
nontrivial possibility is illustrated by  Roberts' example of a
continuum of real solutions in the Newtonian 5-problem with one
negative mass \cite{Roberts}.

The paper is organized as follows. The derivation of the
Albouy-Chenciner equations is shown in Section \ref{AC}. In Section
\ref{HOM} we introduce the homotopy continuation method and several
related software packages. The computation results are described in
Section \ref{computing}. In Section \ref{results} all the central
configurations of the five-body problem with equal masses are
presented.

\section{Albouy-Chenciner Equations\label{AC}}

The Albouy-Chenciner equations are algebraic equations satisfied by
the mutual distances $r_{ij}$ of every central configuration \cite
{Albouy1998,Hampton06}.~ For convenience of the reader, in this section, we present a beautiful derivation of the Albouy-Chenciner equations due to Hampton and Moeckel \cite{Hampton06}. 

Multiplying the $j$-th equation of (\ref{cc}) by $m_j$ and summing
gives $mq_G=\sum_{j=1}^n m_j q_j$, where $m=\sum_{j=1}^n m_j$. Then,
after setting $\lambda=m\lambda^{\prime}$, the central configuration
equations become:
\begin{equation}
\sum_{i=1}^n m_i S_{ij}(q_i-q_j)=0~, \quad 1\leq j\leq n \text{~,}
\end{equation}
where
\begin{equation}
S_{ij}= \frac{1}{r_{ij}^3}+\lambda'\quad (i\neq j), \quad S_{ii}=0
\text{~.} \label{eqS}
\end{equation}

Introducing a $d\times n$ configuration matrix $Q$ whose columns are the
position vectors $q_i$ we can write equation (\ref{cc}) as
\begin{equation}
QA=0 \text{~,} \label{eqQA}
\end{equation}
where $A$ is the $n\times n$ matrix with entries:
\begin{equation}
A_{ij}= m_iS_{ij} \quad (i\neq j) \quad\quad A_{jj}=-\sum_{i\neq
j}A_{ij}\text{~.} \label{eqA1}
\end{equation}

The matrix $Q$ can be viewed as representing a linear map from a
space of dimension $n$ ($n$ being the number of bodies) to the
physical space in ${\mathbb R}^d$ where the points are located.

The most important point of the derivation is to replace $Q$ by
some quantity which is invariant under rotations and translations of
the position vectors $q_i$ in ${\mathbb R}^d$. Of course any such
quantity can be expressed as a function of the mutual distances
$r_{ij}$.

Translation of all the positions $q_i$ by a vector $u\in{\mathbb
R}^d$ transforms $Q$ to $Q+uL$ where $L$ is the $1\times n$ vector
whose components are all $1$. Thus translation invariance can be
achieved by restricting the linear map defined by $Q$ to the plane
$P=\{v\in{\mathbb R}^n: Lv=v_1+\ldots+v_n=0\}$. Since the sum of the
elements of each column of $A$ is zero it follows that $A:{\mathbb
R}^n\rightarrow P$ and it restricts to $A:P\rightarrow P$.~ So $QA$
can be viewed as a linear map of $P$ into ${\mathbb R}^d.$

Rotation invariance is obtained by passing to Gram matrices. For any
configuration matrix $Q$, the Gram matrix $G=Q^{t}Q$ is the $n\times
n$ matrix whose entries are the Euclidean inner products $q_i\cdot
q_j$.~ $G$ is obviously rotation invariant. To maintain translation
invariance, one can view $G$ as representing a symmetric bilinear
form $\beta(v,w)=v^{t}Gw$ on $P$. The form $\beta$ on $P$ determines
and is determined by the mutual distances $r_{ij}$. To see this,
note that for any constant $k_i$, adding the vector $k_iL$ to the
row $i$ of $G$ and the vector $k_iL^t$ to the column $i$ produces a
new matrix representing $\beta$. Choosing $k_i=-\frac 1 2 |q_i|^2$
shows that $\beta$ is represented by the matrix $B$ whose entries
are
\begin{equation}
B_{ij}=q_i\cdot q_j-\frac 1 2|q_i|^2-\frac 1 2|q_j|^2=-\frac 1 2 r_{ij}^2.
\label{eqB}
\end{equation}

Multiplying both sides of (\ref{eqQA}) by $Q^t$ gives $GA=0$. The matrix $GA$
can be viewed as representing a (non-symmetric) bilinear form on $P$, in
which case it is permissible to replace $G$ by $B$. Taking the symmetric
part gives the Albouy-Chenciner equations for central configurations:
\begin{equation}
BA+A^tB=0.  \label{eqBA}
\end{equation}

Let $e_i$ denote the standard basis vectors in ${\mathbb R}^n$ and
define $e_{ij}=e_i-e_j$. Then (\ref{eqBA}) is equivalent to the
equations
\begin{equation}
e_{ij}^t(BA+A^tB)e_{ij}=0 \quad 1\leq i<j\leq n.  \label{eqe}
\end{equation}

To see this, let $\gamma(v,w)=v^tCw$ be the symmetric bilinear form
on $P$ associated to the matrix $C=BA+A^tB$. Then (\ref{eqe}) means
that $\gamma(e_{ij},e_{ij})=0$ for $1\leq i<j\leq n$. To show that
$\gamma=0$ it suffices to show that $\gamma$ vanishes on the basis
$e_{1i}$, $2\leq i\leq n$ of $P$. By the polarization identity
\[
2\gamma(e_{1i},e_{1j})=\gamma(e_{ij},e_{ij})-\gamma(e_{1i},e_{1i})-
\gamma(e_{1j},e_{1j}) \text{,}
\]
this follows from (\ref{eqe}).

Equations (\ref{eqe}) provide ${\binom{ n }{2 }}$ constraints on the
${\binom{ n }{2 }}$ mutual distances $r_{ij}$ of a central
configuration. Conversely, it can be shown that if the quantities
$r_{ij}$ are mutual distances of some configuration in ${\mathbb
R}^d$, and if they satisfy (\ref{eqe}), then the configuration is
central. Note that these equations determine the central
configurations in all dimensions at once.

To find the equation explicitly, note that
\[
\gamma(e_{ij},e_{ij})=2e^t_{ij}BAe_{ij}=2((BA)_{ii}+(BA)_{jj}-(BA)_{ij}-(BA)_{ji})\text{,}
\]
where $(BA)_{ij}$ denotes the entries of the matrix $BA$. From (\ref{eqA1})
and (\ref{eqB}) we find
\begin{equation}
\sum_{k=1}^n m_k
[S_{ik}(r_{jk}^2-r_{ik}^2-r_{ij}^2)+S_{jk}(r_{ik}^2-r_{jk}^2-r_{ij}^2)]=0
\label{eqAC}
\end{equation}
for $1\leq i< j\leq n$, where $S_{ik}$ and $S_{jk}$ are given by (\ref{eqS}).

Finally, to eliminate the dilation symmetry, we choose
$\lambda'=-1$.


\section{The homotopy continuation method\label{HOM}}

Let $P(x)=0$ be a system of ~$n$~ polynomial equations in ~$n$~
unknowns.~ Denoting ~$P(x)=\left( p_{1}(x),\dots ,p_{n}(x)\right)
\,$~ and ~$\,x=\left( x_{1},\dots ,x_{n}\right)$,~ we want to find
all the isolated solutions of
\[
P(x)=\left\{
\begin{array}{c}
p_{1}\left( x_{1},\dots ,x_{n}\right) =0 \\
\vdots \\
p_{n}\left( x_{1},\dots ,x_{n}\right) =0
\end{array}
\right.
\]
in $\bdC^{n}$. ~The classical homotopy continuation method
\cite{Lihand} for solving ~$P(x)=0$~ is to find a trivial system
$Q(x)=\left( q_{1}(x),\dots ,q_{n}(x)\right) $ and then follow the
solution curves in the real variable ~$t$~ from ~$t=0$~ to ~$t=1$~
which make up the solution set of
\[
H(x,t)=(1-t)c Q(x)+tP(x)=0\text{ \ \ \ with generic \ }c \in \bdC
\backslash \{0\}\text{.}
\]
More precisely, all isolated solutions of ~$P(x)=0$~ can be found if
the system ~$Q(x)=0$,~ known as the start system, is chosen properly
to satisfy the following three properties:
\begin{itemize}
\item  Property 0. The solutions of the start system ~$Q(x)=0$~ are known;

\item  Property 1. The solution set of ~$H(x,t)=0$~ for ~$0\leq t\leq
1$~
consists of a finite number of smooth paths, and each of them can be
parametrized by $t$ in $[0,1)$;

\item  Property 2. Every isolated solution of ~$H(x,1)=P(x)=0$~ can be reached
by some path originating at ~$t=0$,~ that is, the path starts from a
solution of the start system ~$H(x,0)=Q(x)=0$.
\end{itemize}
\begin{figure}[ht]
\input epsf \centerline{\epsfxsize=3.6in
\epsfysize=2.8in \epsfbox{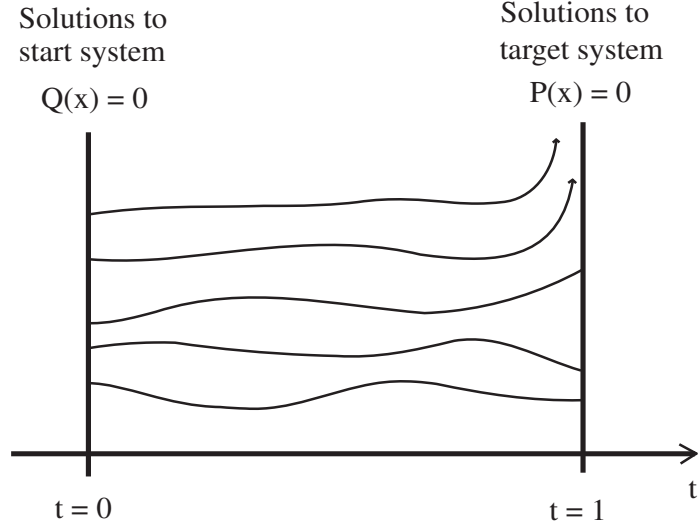}} \vspace{-3mm} \caption{The
solution set of classical homotopy.}\label{curves}
\end{figure}
A typical choice of a start system ~$Q(x)=0$~ satisfying Property
0-2 is
\[
Q(x)=\left\{
\begin{array}{c}
q_{1}\left( x_{1},\dots ,x_{n}\right) =a_{1}x_{1}^{d_{1}}-b_{1} \\
\vdots  \\
q_{n}\left( x_{1},\dots ,x_{n}\right) =a_{n}x_{n}^{d_{n}}-b_{n}
\end{array}
\right.
\]
where ~$d_{1},\cdots ,d_{n}$~ are the degrees of polynomials
~$p_{1}(x),\dots ,p_{n}(x)$~ respectively, and $a_{j}$, $b_{j}$,
$j=1,\cdots ,n$ are random complex numbers \cite{Li1,Li2,Lihand}.
The solutions of such start system $Q(x)=0$ can be explicitly
obtained and the total number of solutions is $d=d_{1}\times
d_{2}\times \cdots \times d_{n}$, which is known as the \emph{total
degree} or the B\'{e}zout number of the original polynomial system
$P(x)=0$. ~We may then find all the isolated solutions of ~$P(x)=0$~
by following the total degree number of paths originated from
solutions of the start system ~$Q(x)=0$.~ But, a great majority of
the polynomial systems arising in applications have fewer than, and
in some cases only a small fraction of, ~$d=d_{1}\times \cdots
\times d_{n}$~ isolated zeros. ~We call such a system
\emph{deficient}. In this case, many of the ~$d_{1}\times \cdots
\times d_{n}$~ paths will diverge to infinity as ~$t\rightarrow
1$~(see Figure \ref{curves}), and those paths become extraneous,
causing highly wasteful computations.

In the middle of 1990's, a major computational breakthrough emerged
in solving deficient polynomial systems. ~The new method, called the
\emph{polyhedral homotopy} continuation method \cite{HS}, takes a
great advantage of the combinatorial root count, called \emph{mixed
volume}, in the Bernshte\'in's theorem \cite{Bernshtein}, which
generally provides a much tighter bound for the number of isolated
zeros of a polynomial system in the algebraic torus $(\bdC^{\ast
})^{n}=(\bdC\backslash \{0\})^{n}$. ~When this method is employed,
the number of homotopy paths that need to be traced agrees with the
mixed volume of the polynomial system. ~Therefore, when the mixed
volume of a polynomial system is far less than its total degree, the
method will greatly reduce the extraneous paths and thereby
considerably limit the wasteful computations.

However, this method involves a sometimes costly computation - the
\emph{mixed cell} computation. This mixed cell computation can
become very costly for large polynomial systems.
~In 2005, a software package, MixedVol
\cite{Gao3}, emerged which led the existing codes for the mixed
volume computation by a great margin in speed.~ However, soon after
MixedVol was published, Mizutani, Takeda, and 
Kojima \cite{Mizutani1} developed a more efficient algorithm which
considerably outperformed MixedVol. Most
recently  Lee and Li \cite{Le} embedded the novel idea of
\emph{dynamic enumeration} of mixed cells of Mizutani, Takeda, and Kojima \cite{Mizutani1}  into
MixedVol and a new code, MixedVol-2.0, was
produced which regained the lead by a substantial margin.

The software package HOM4PS \cite{Gao3}, developed over the years by
the group led by T. Y. Li at Michigan State University, implemented
the polyhedral homotopy continuation method for solving polynomial
systems. ~The code was widely considered as the most efficient
polynomial systems solver. Recently many new curves tracing
techniques and mixed cell computation algorithms, such as
MixedVol-2.0 mentioned above, have been included in HOM4PS-2.0. By
incorporating these new algorithms HOM4PS-2.0 is (as reported in
\cite{HOM4PS2}), in general, at least one order of magnitude faster
than  HOM4PS, as well as from one to three orders of magnitude
faster than PHCpack \cite{PHCpack}, PHoM \cite{PHOM} and Bertini
\cite{Bertini}. ~Consequently, some large size polynomial systems
can be solved in a reasonable number of  hours. ~However, as the
size of the polynomial system becomes larger, we need more computing
resources to solve the system more efficiently. ~A natural way to
attain more computing resources is to parallelize the homotopy
continuation method which seems to be naturally parallel in the
sense that each isolated zero is computed independently of the
others. ~The parallel version of HOM4PS-2.0, HOM4PS-2.0para, has
also been developed. Its excellent scalability in numerical results
are reported in \cite{HOM4PS2para}.


\section{Computing central configurations\label{computing}}


The Albouy-Chenciner equations for the five-body problem with equal
masses form a polynomial system consisting of $20$ equations in $20$
variables.
For finding all the isolated solutions in complex field, the polyhedral
homotopy method needs to trace ~$439,690,761$~ curves.
In order to trace so many curves efficiently, we use the subroutines
in the MPI library (message passing interface \cite{MPI}) to
distribute data over multiple CPUs for parallel computation and we
use the kernel of HOM4PS-2.0 to construct polyhedral homotopies and
trace curves.
Employing $32$ Itanium2 1.6 GHz CPUs, the computation is completed
in $140$ hours.
As a result we obtain  ~$101,062,826$~ solutions in the complex field, ~$8775$~ of which are real solutions.
Among those real solutions only 258 satisfy the physical condition
(i.e. $r_{ij}>0$ for all $1\leq i<j\leq 5$).
%
%
We assume that any numerical solution with imaginary parts less than
the threshold $\theta = 10^{-7}$ corresponds to a real solution of
the Albouy-Chenciner equations.
This is reasonable since the set of real solutions remains the same
when the threshold $\theta$ is chosen between $10^{-13}$ and
$10^{-3}$.
In addition, the residuals of the physical solutions are less than
$2\times 10^{-14}$ and their condition numbers are at most
$3.8\times 10^2$, which show that these solutions are numerically
reliable.

To guarantee that in a small neighborhood of each numerical solution
there is a unique exact physical solution, the interval Krawczyk
method \cite{Kraw,Moore77} is applied for verification.
The method is based on the fact:~ For a smooth function $F:\bdR
^{n}\rightarrow \bdR^{n}$ and a point $\mathbf{x}\in \bdR^{n}$, let
~$[\;\mathbf{x}\;]_{r}\subset \bdR^{n}$~ be the interval set
centered at $\mathbf{x}$ with radius $r>0$.~ Namely,
\[
[\;\mathbf{x}\;]_{r}~=~\left\{ \mathbf{y}\in \bdR^{n}:\left\| \mathbf{y}-%
\mathbf{x}\right\| _{\infty }\leq r\right\} \text{,}
\]
where $\left\| \mathbf{\cdot }\right\| _{\infty }$ is the infinity
norm. Assume the derivative of $F$ at $\mathbf{x}$, denoted by
~$DF(\mathbf{x})$, ~is nonsingular, the Krawczyk set of $F$
associated with ~$[\;\mathbf{x}\;]_{r}$~ is defined as
\[
K(F,[\;\mathbf{x}\;]_{r})~=~\mathbf{x}-DF(\mathbf{x})^{-1}F(\mathbf{x})+\left[
I-DF(\mathbf{x})^{-1}DF([\;\mathbf{x}\;]_{r})\right] ([\;\mathbf{x}\;]_{r}-\mathbf{x}%
)\text{.}
\]
If the Krawczyk set is contained in the interior of
~$[\;\mathbf{x}\;]_{r}$,~ then there exists a unique zero of $F$
in ~$[\;\mathbf{x}\;]_{r}$.

The task of verification is implemented by using
the interval arithmetic in INTLAB (INTerval LABoratory) \cite{Rump},
developed by Siegfried M. Rump at Hamburg University of Technology,
Germany.
~In this implementation each numerical solution $\mathbf{x}$ is
taken as the midpoint of the interval set ~$[\;\mathbf{x}\;]_{r}$~
with radius $r=10^{-8}$.
The numerical solutions obtained by running HOM4PS-2.0 and the
results of verification by using INTLAB are available for download
at {\tt http://hom4ps.math.msu.edu/Nbody.htm}~.


\section{Central Configurations of the 5 body problem\label{results}}

In this section we present all the central configurations of the
five-body problem with equal masses classified according to their
dimensions. The dimension of the configuration was deduced by using
Cayley-Menger determinants \cite{Sommerville}. For instance the
Cayley-Menger determinant for the volume $V$ of the four-dimensional
simplex is
\[
-9216V^2=\left|\begin{array}{cccccc}
0&1&1&1&1&1\\
1&0&r_{12}^2&r_{13}^2&r_{14}^2&r_{15}^2\\
1&r_{12}^2&0&r_{23}^2&r_{24}^2&r_{25}^2\\
1&r_{13}^2&r_{23}^2&0&r_{34}^2&r_{35}^2\\
1&r_{14}^2&r_{24}^2&r_{34}^2&0&r_{45}^2\\
1&r_{15}^2&r_{25}^2&r_{35}^2&r_{45}^2&0
\end{array}\right|.
\]
 Configurations  of dimension three or less were found by setting the determinant above to zero.
 Configurations of dimension two or less were found by setting to zero the determinant above plus
the following  determinants
\[
\left|\begin{array}{cccccc}
0&1&1&1&1\\
1&0&r_{12}^2&r_{13}^2&r_{14}^2\\
1&r_{12}^2&0&r_{23}^2&r_{24}^2\\
1&r_{13}^2&r_{23}^2&0&r_{34}^2\\
1&r_{14}^2&r_{24}^2&r_{34}^2&0\\
\end{array}\right|,~~
\left|\begin{array}{cccccc}
0&1&1&1&1\\
1&0&r_{12}^2&r_{13}^2&r_{15}^2\\
1&r_{12}^2&0&r_{23}^2&r_{25}^2\\
1&r_{13}^2&r_{23}^2&0&r_{35}^2\\
1&r_{14}^2&r_{24}^2&r_{35}^2&0\\
\end{array}\right|,~~
\left|\begin{array}{cccccc}
0&1&1&1&1\\
1&0&r_{23}^2&r_{24}^2&r_{25}^2\\
1&r_{23}^2&0&r_{34}^2&r_{35}^2\\
1&r_{24}^2&r_{34}^2&0&r_{45}^2\\
1&r_{25}^2&r_{35}^2&r_{45}^2&0\\
\end{array}\right|,
\]
 where, for example, the first of the three determinants above is the
Cayley-Menger determinants for the tetrahedron formed by the masses
$m_1, m_2, m_3$ and $m_4$.
One-dimensional configurations were obtained by setting to zero, in
addition to the four Cayley-Menger determinants above, the ten
Cayley-Menger determinants corresponding to all the possible
triangles formed by a subset of three of the five masses. One of the
ten determinants  is
\[-16A^2=\left|\begin{array}{cccccc}
0&1&1&1\\
1&0&r_{12}^2&r_{13}^2\\
1&r_{12}^2&0&r_{23}^2\\
1&r_{13}^2&r_{23}^2&0\\
\end{array}\right|,
\]
where $A$ is the area for a plane triangle with side lengths $r_{12},r_{13}$ and $r_{23}$.

\begin{figure}[t]
\centering
\subfloat[][]{\resizebox{!}{5cm}{\includegraphics{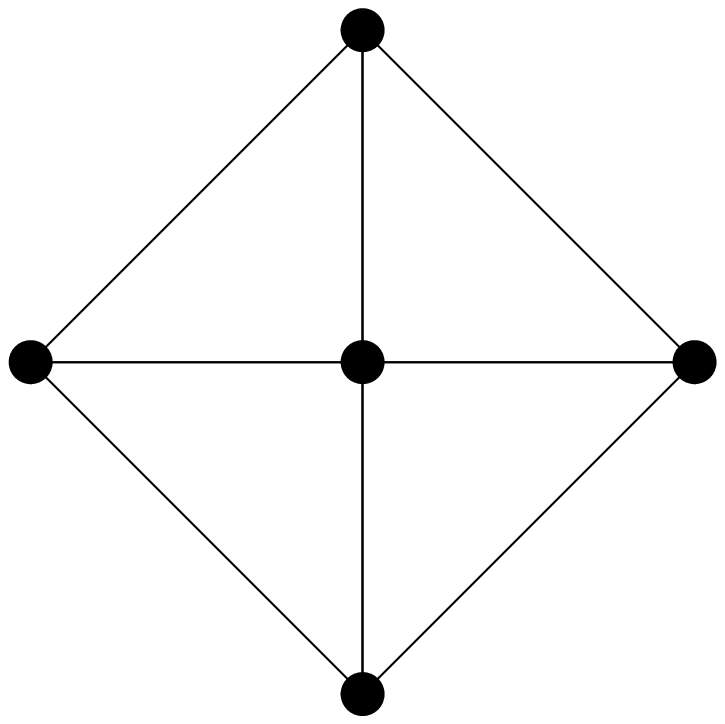}}}
\qquad\quad\quad \subfloat[][]{\rotatebox{90}{\resizebox{!}{5cm}{%
\includegraphics{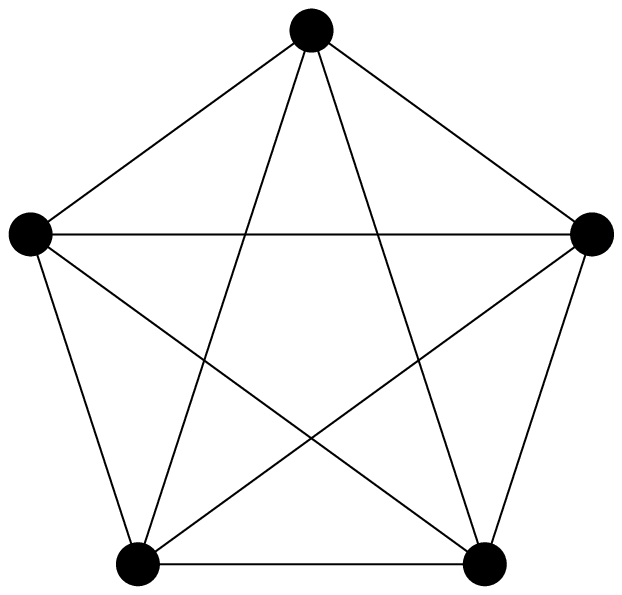}}}} 
\par
\par
\subfloat[][]{\rotatebox{270}{\resizebox{!}{5cm}{\includegraphics
   {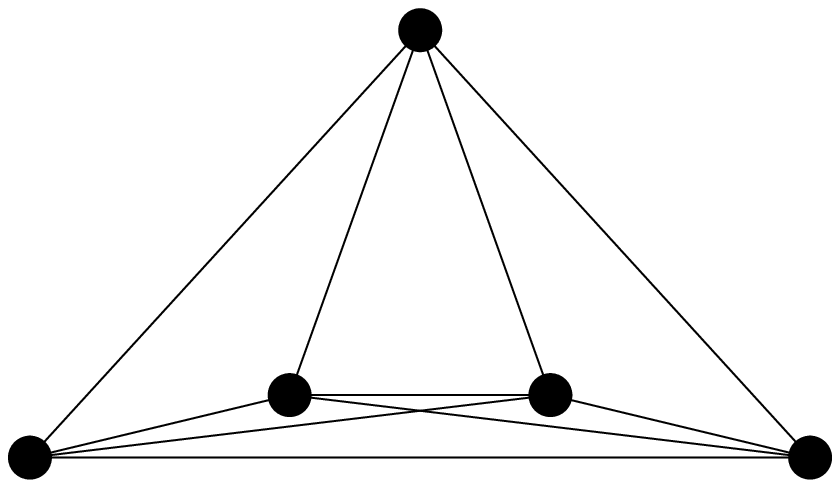}}} } \qquad
\subfloat[][]{\rotatebox{270}{\resizebox{!}{5cm}{\includegraphics{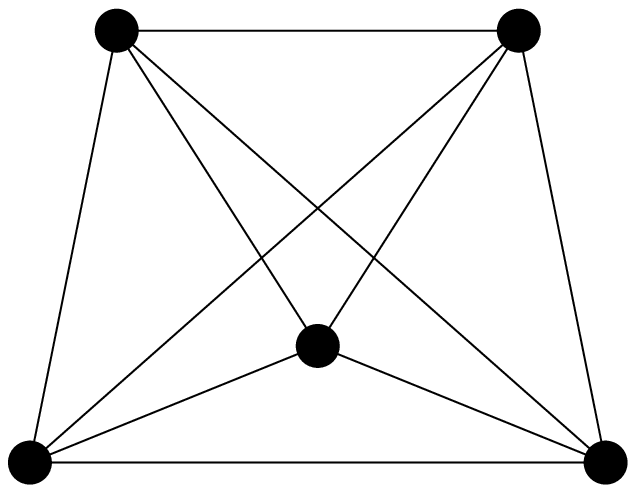}}}
     }
\caption{Planar central configurations of the five body problem with equal
masses. }
\label{planarcc}
\end{figure}

\subsection{Collinear and four-dimensional central configurations}

A complete count of the collinear central configurations has been known for
almost one century. In 1910 Moulton \cite{Moulton} proved that there is a
unique central configuration corresponding to each possible ordering of $n$
masses along a line, and thus there are $n!/2$ collinear central
configurations. In 1970 Smale used tools of the theory of dynamical systems
to give a modern proof of Moulton's result.

Our computations show, in agreement with Moulton's result, that there are $%
60=5!/2$ solutions of the Albouy-Chenciner equations.

The four-dimensional central configurations of five bodies are also
well understood. Saari \cite{Saari} proved that the regular $n-1$
dimensional simplex is a central configuration of $n$ bodies for any
value of the masses. In particular the case $n=4$ has been well
known over a century \cite{Lehmann}. The fact that the tetrahedron
is the unique spatial central configuration of four bodies was
proved by Pizzetti \cite{Pizzetti}. We find one 4-dimensional
solution of the Albouy-Chenciner equations that, as we expected, is
a regular 4-dimensional simplex.

\subsection{Planar Central Configurations}

There are very few results concerning the planar five-body problem.
Part of the work on the planar five body problem dates back to more
than 50 years ago and is due to Williams \cite{Williams38},\cite{Williams53}.
 More recently, in 2005, Hampton \cite{Hampton05} found
new examples of (symmetric) central configurations in the planar
five-body problem. Llibre and Mello \cite {Llibre08} and Llibre, Mello, and Perez-Chavela \cite{LMP} found some further central configurations
using the same techniques employed by Hampton. Not much more is
known in the planar five-body problem

It has been known for a long time that, in the equal mass case, the regular
pentagon and the square with one mass at each of its vertices and with the
fifth mass at its center are central configurations \cite{Williams38}.

In the case of equal masses we find 147 planar solutions of the
Albouy-Chenciner equations. Of these, 15 form a square with a
particle in the middle (Figure \ref{planarcc} (a)), 12 form a
regular pentagon (Figure \ref{planarcc} (b)), 60 form an isosceles
triangle with the remaining particles in the convex hull of the
first three (Figure \ref{planarcc} (c)), and 60 form an isosceles
trapezoid with a mass in the convex hull of the first four (Figure
\ref{planarcc} (d)). The configurations of Figure \ref {planarcc}
(c) and Figure \ref{planarcc} (d) have been found numerically in
1989 by  Moeckel  \cite{Moeckel89} and rediscovered by  Ferrario
\cite{Ferrario}. Note that  in \cite{Williams38} Williams
erroneously states that a concave pentagon with three masses at the
vertices of a pentagon and the remaining two in its interior cannot
form central configurations. Our results (see figure Figure \ref
{planarcc} (c)) show  that such statement is incorrect.

It is interesting to remark that the number of solutions obtained
can be explained using their symmetry. In the case of the
configuration of Figure (\ref{planarcc}(a)) the four masses at the
vertices of a square and one at the center are subject to an action
of the group $S_5$ (symmetric group on 5 letters) permuting the
masses and hence the set of mutual distances $r_{ij}$.  But, the set
of mutual distances $R$ for any particular numbering of the masses
has an {\it isotropy subgroup} that is equal to the symmetries of
the square (the dihedral group $D_4$ of order $8$).  Hence the set
of $S_5$-orbits of configurations ( that is equal to the number of
solutions of the Albouy-Chenciner equations) of this type should
have $120/8 = 15$ elements. Similarly, the pentagon configurations
in Figure  (\ref{planarcc}(a)) have the dihedral group $D_5$ as
isotropy subgroup and there are $120/10 = 12$ orbits. The
configurations in Figure (\ref{planarcc} (a)) and (\ref{planarcc}
(b)) have $C_2$ (i.e. the cyclic group of order 2)    as isotropy
subgroup and, in this case,  there are $120/2=60$ orbits.

Another interesting fact is that, as in the four-body problem with equal
masses \cite{Albouy95}, all the planar central configuration have an axis of
symmetry that contains at least one mass.

\subsection{Spatial Central Configurations}

\begin{figure}[t]
\centering
\par
\subfloat[][]{\resizebox{!}{7.cm}{\includegraphics{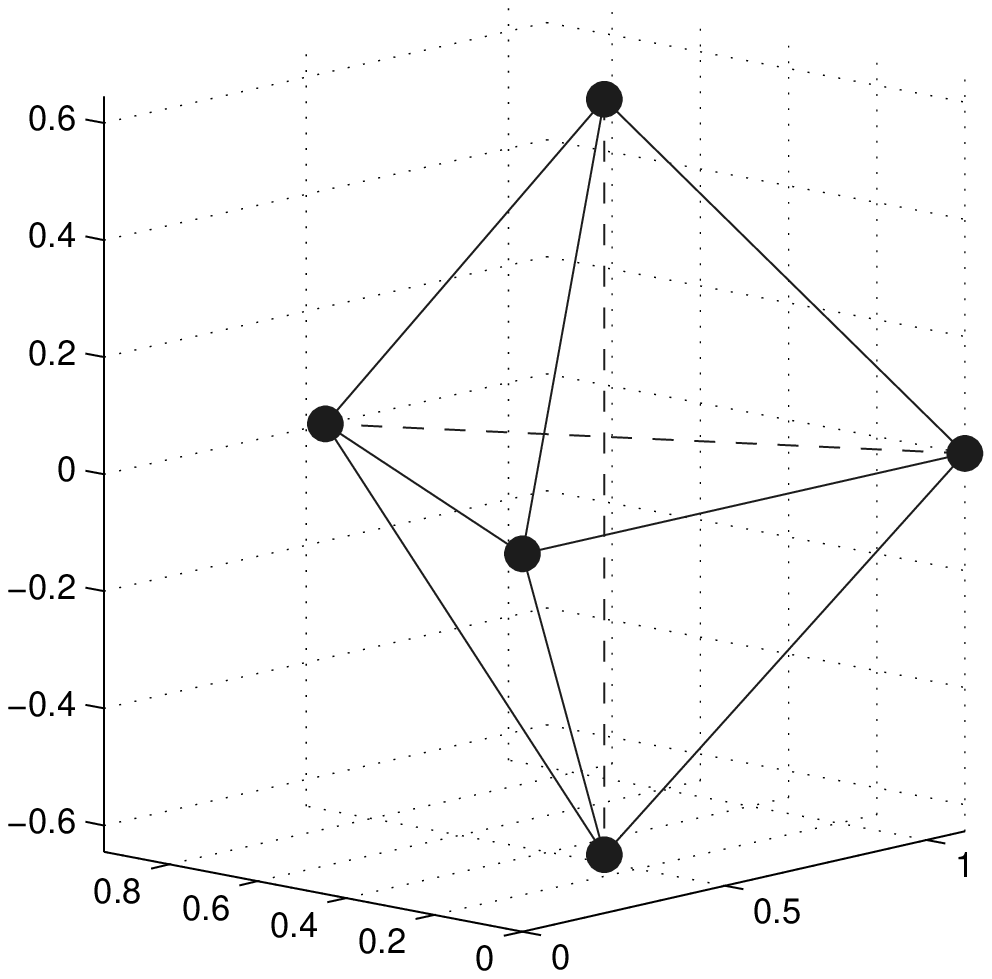}}} %
\subfloat[][]{\resizebox{!}{7.cm}{\includegraphics{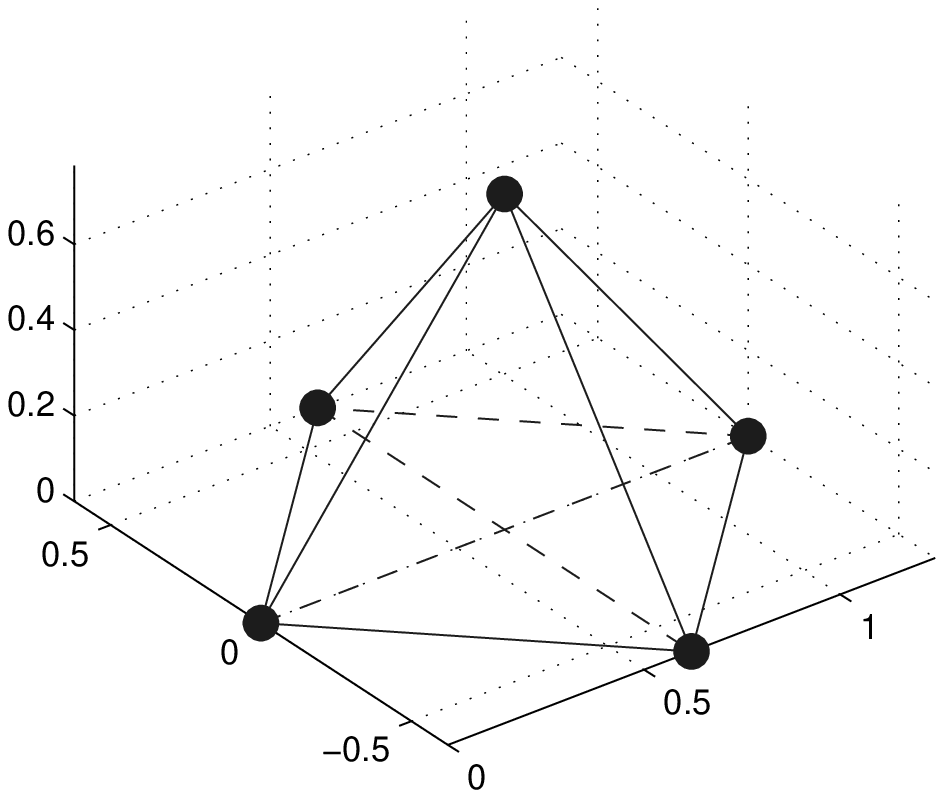}}}
\par
\caption{Convex spatial central configurations of five bodies with equal
masses.}
\label{fig3Dconvex}
\end{figure}

\begin{figure}[t]
\centering
\par
\subfloat[][]{\resizebox{!}{7.cm}{\includegraphics{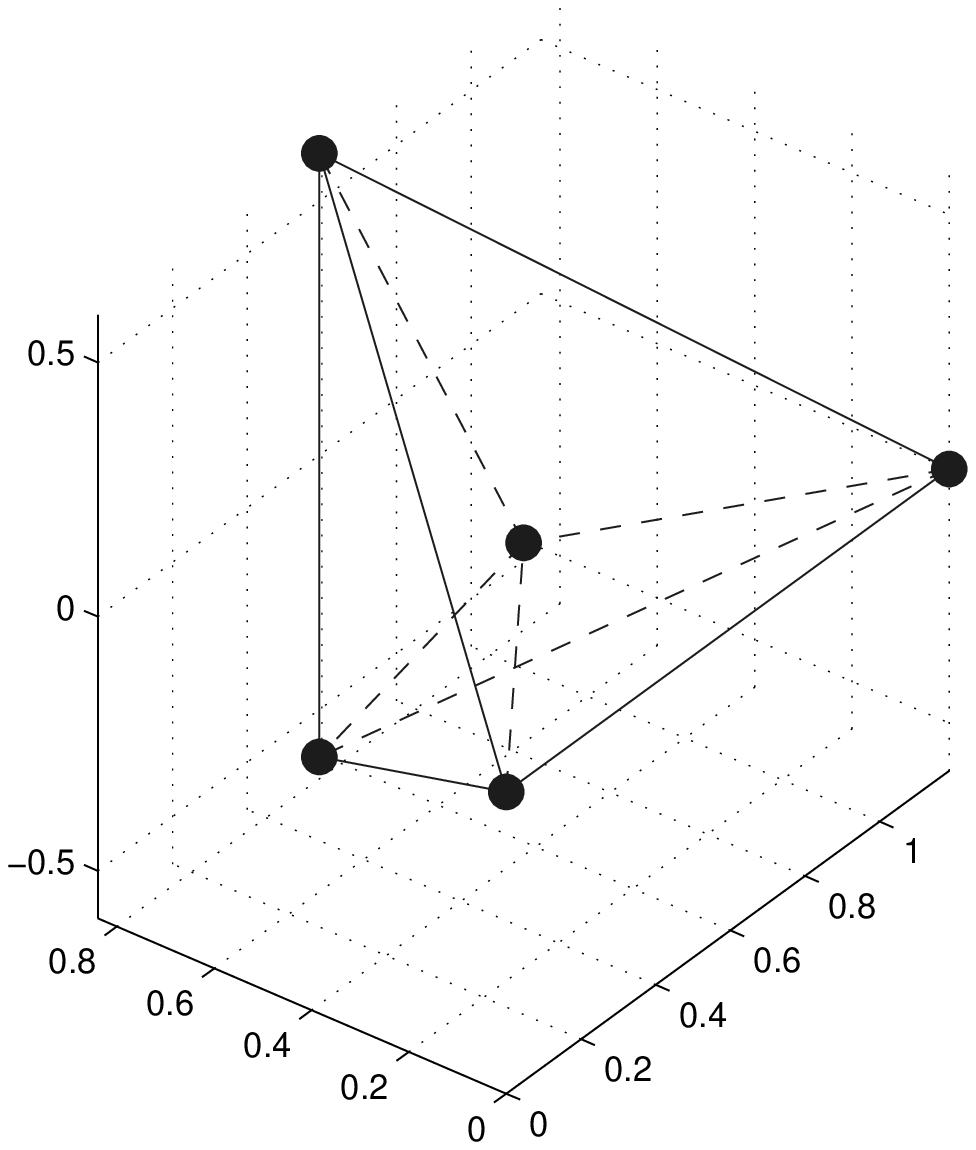}}} %
\subfloat[][]{\resizebox{!}{7.cm}{\includegraphics{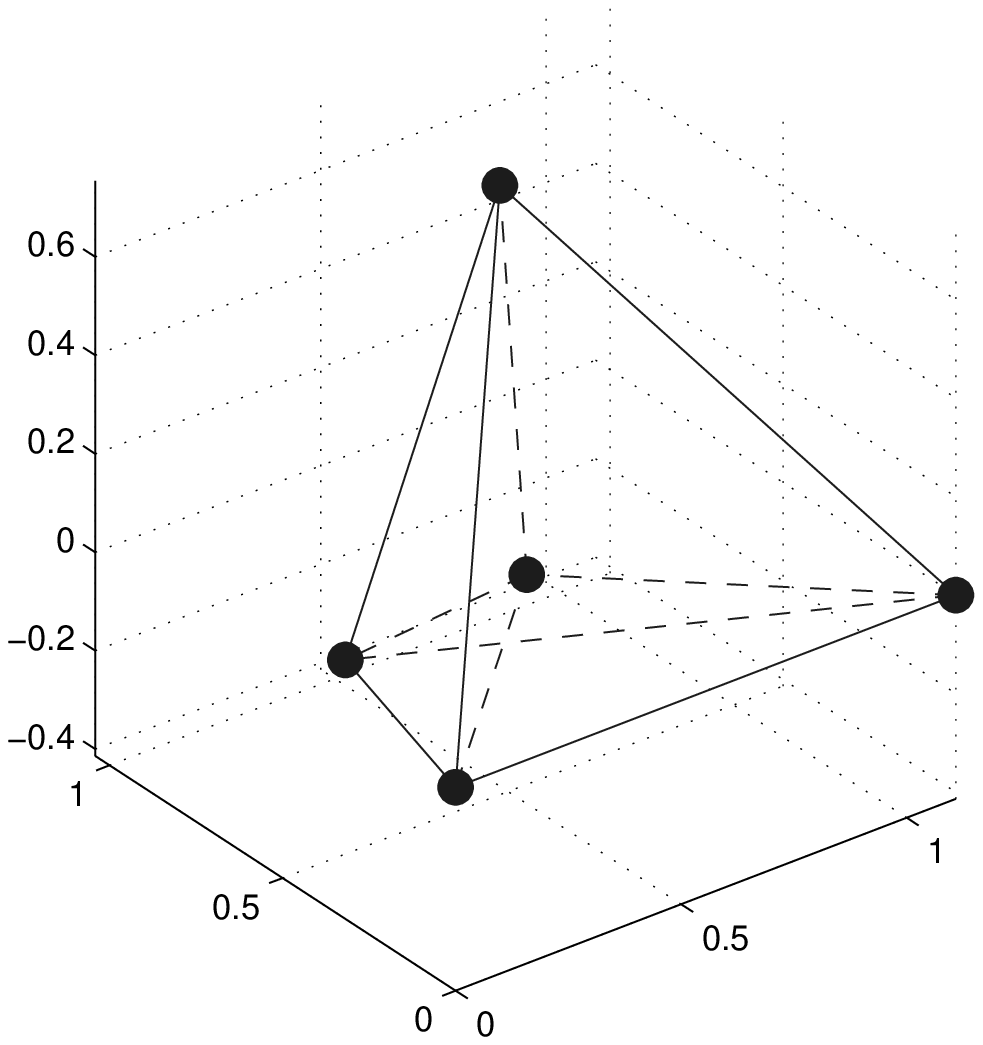}}}
\caption{Concave spatial central configurations of five bodies with equal
masses.}
\label{fig3Dconcave}
\end{figure}

In 1908 Brehm \cite{Brehm} found some symmetric spatial central
configurations of five bodies. Almost one century later Kotsireas
and Lazard \cite{Kotsireas} used linear algebra and Gr\"obner bases
to classify symmetric spatial central configurations of five bodies
with equal masses. They found four distinct classes of central
configurations. Kotsireas and Lazard conjectured that their list of
central configurations was complete. Xia proved, for general masses,
that there is at least one convex central configuration that is a
minimum of the configurational measure \cite{Xia}. Moeckel used
Morse theory, under the assumption of nondegenerate critical points,
to find 6 more central configurations of saddle type \cite{Moeckel}.

We find 50 spatial solutions of the Albouy-Chenciner equations. These
correspond to four distinct classes of central configurations (two convex
and two concave) that match the ones of Kotsireas and Lazard, thus proving
their conjecture to be correct.

The first class has 10 solutions. These consist of an equilateral
triangle and two symmetric points on the axis through the barycenter
of the triangle that is orthogonal to it (Fig.
\ref{fig3Dconvex}(a)). The second one contains 15 solutions. These
are pyramidal central configurations with a square base (Fig.
\ref{fig3Dconvex}(b)). The third class consists of 5 solutions.
These consist of a regular tetrahedron with a point in its
barycenter (Fig. \ref{fig3Dconcave}(a)). The last class contains 20
solutions. These consist of an equilateral triangle and two points
lying on the axis through the center of the triangle (Fig.
\ref{fig3Dconcave}(b)).

As in the planar case  the number of central configurations can be
explained using their symmetry. In the case of the configuration of
Figure (\ref{fig3Dconvex}(a)) the masses  are subject to an action
of the group $S_5$.  But, the set of mutual distances  has the group
$D_3\times C_2$  (of order 12) as  isotropy subgroup .  Hence the
set of $S_5$-orbits of configurations  of this type should have
$120/12 = 10$ elements and thus there are $10$ solutions of the
Albouy Chenciner equations. The  configuration in Figure
(\ref{fig3Dconvex}(b)) has the dihedral group $D_4$ as isotropy
subgroup and there are $120/8 = 15$ solutions. The configuration in
Figure (\ref{fig3Dconcave} (a))  has the tetrahedral group $T_d$
(that has order 24)  as isotropy subgroup and hence there are
$120/24=5$ solutions.

The configuration in Figure (\ref{fig3Dconcave} (b)) has the
dihedral group $D_3$  as isotropy subgroup and, in this  case, there
are $120/6=20$ solutions.

According to Moeckel \cite{Moeckel} the convex central configurations would
fit with the Morse theory as follows: 6 saddles plus 9 more =15, 1 minimum
plus 9 more = 10.

\section*{Acknowledgments}

This work was made possible by the facilities of the Shared
Hierarchical Academic Research Computing Network
(SHARCNET:www.sharcnet.ca) and by INTLAB (INTerval LABoratory) a
Matlab toolbox for self-validating algorithms, developed by
Siegfried M. Rump at Hamburg University of Technology, Germany. The
authors thank Chih-Hsiung Tsai for his  assistance in collecting all
the complex solutions of the Albouy-Chenciner equations, and Alain
Albouy for bringing to our attention unpublished numerical
computations by Moekel and Ferrario \cite{Ferrario}.
Manuele Santoprete was supported by a NSERC Discovery grant.

\end{document}